# Near-UV Single-Pixel Imaging with All-Inorganic Lead-Free Perovskite

Jiyun Kim[1, 2*], Xinyang Yu[1, 2], Zijian Feng[3], Chun-Ho Lin[3], Dewei Chu[3], Igor Aharonovich[1, 2], Chaohao Chen[1,2,4*]

[1]School of Mathematical and Physical Sciences, University of Technology Sydney, Ultimo, New South Wales 2007, Australia

[2]ARC center of Excellence for Transformative Meta-Optical Systems, University of Technology Sydney, Ultimo, New South Wales 2007, Australia

[3]School of Materials Science and Engineering, University of New South Wales, Sydney, NSW 2052, Australia

[4]School of Biomedical Engineering, University of Technology Sydney, Ultimo, New South Wales 2007, Australia

*Correspondence to: chloe.kim@uts.edu.au; chaohao.chen@uts.edu.au

**Abstract**

*Single-pixel imaging (SPI) is a powerful computational imaging technology that reconstructs spatial information from sequentially encoded optoelectrical signals without pixelated detector arrays. Solution-processible metal halide perovskites are promising photoactive candidates for SPI, but the toxicity of lead-based compositions remains a critical barrier to practical development. Here, we demonstrate one-step fabrication of low-dimensional, lead-free* $K_2CuBr_3$ *thin film as near-UV photoactive channels for single-pixel imaging. By systematic antisolvent engineering, compact and uniform* $K_2CuBr_3$ *films are obtained and integrated into planar photoconductors devices. The resulting photodetectors exhibit stale photoswitching under 405 nm illumination, low dark current on the order of* $10^{-10}$ *A, with fast response and recovery time 38.82 and 61.94 μs, respectively. Integrated into an SPI configuration, the* $K_2CuBr_3$ *photoconductor successfully reconstructs near-UV images, with the signal-to-noise ratio improving from 16.4 to 31.7 dB as the illumination irradiance increases. This work highlights solution-processed lead-free copper halides as promising photoactive materials for compact, non-toxic and cost-effective UV computational imaging systems.*

*Key words: Single-pixel imaging, lead-free perovskite,* $K_2CuBr_3$*, photodetector*

## Introduction

Ultraviolet (UV) and near-UV imaging is important for environmental monitoring, material inspection and optical security.[1,2] Conventional UV imaging technologies typically rely on dense focal-plane pixelated detector arrays, which requires complex device fabrication and readout circuitry.[3] These challenges become more pronounced in the UV range, where detector materials must combine strong short-wavelength absorption, low dark current, fast temporal response and operational stability.[4,5] Developing compact and low-cost UV imaging architectures therefore remains highly desirable.

Single-pixel imaging (SPI) provides an alternative route to optical imaging by replacing pixelated detector arrays with a single photosensitive element and computational reconstruction.[6,7] In a typical SPI system, spatial information is recoded into a sequence of structured illumination patterns, and the integrated detector response is correlated with the known patterns to image reconstruction. This architecture is particularly attractive for spectral regions where high-performance detector arrays are costly or difficult to fabricate. However, the imaging speed and reconstruction fidelity of SPI are strongly limited by the temporal response of the single-pixel detector. To follow high-speed pattern projection from a digital micromirror device, the detector must provide a fast and reproducible photocurrent response within each pattern dwell time. For example, a 100 μs pattern duration corresponds to a 10 kHz structured illumination sequence, requiring detector response times in the sub-100 μs regime.[8,9]

All-inorganic halide perovskites are of great interest as solution-processed photoactive materials, offering large absorption coefficients, tunable bandgap, long carrier diffusion lengths, and low-cost film fabrication.[10-12] These features enable them to address a broad range of imaging applications from X-ray scintillators and hyperspectral cameras to UV and near-infrared sensing. Nevertheless, the toxicity and environmental concerns associated with lead-based compositions remain a major barrier to practical application. Lead-free all-inorganic halides, particularly low-dimensional copper halides, offer an attractive alternative. Their wide bandgaps, strong excitonic effect and structural tunability are intrinsically suitable to UV and deep-UV photodetection.[13,14] In addition, their all-inorganic composition can offer improved thermal stability compared with organic–inorganic hybrid perovskites. Despite these advantages, most lead-free perovskite photodetectors have been evaluated mainly through isolated device metrics, while their integration into computational imaging systems has not been demonstrated.

Herein, we report the one-step fabrication of one-dimensional (1D) lead-free $K_2CuBr_3$ thin film as near-UV photoactive channels for single-pixel imaging. By systematic anti-solvent engineering, we achieved high quality of 1D $K_2CuBr_3$ thin film with strong carrier confinement. Leveraging the $K_2CuBr_3$ film as the photoactive channel in a SPI system, we demonstrate successful near-UV image reconstruction with ultrafast response and recovery time (38.82 and 61.94 μs) across a wide irradiance dynamic range from 0.2 to 200 W/cm$^2$ with improving SNR from 16.4 to 31.7 dB. This work can provide insights of $K_2CuBr_3$ as promising photoactive materials for compact, non-toxic and cost-effective UV computational imaging systems.

## 2. Results and Discussion

The solution-processed fabrication of $K_2CuBr_3$ is illustrated in Figure 1a (detailed in Methods). In the context of growing $K_2CuBr_3$ perovskite films, dimethyl sulfoxide (DMSO), used as the primary solvent, dissolves precursors efficiently due to its high polarity and strong coordination with metal cations, forming stable intermediate complexes that regulate nucleation and crystallization.[15] However, antisolvents selection process is significant for perovskite growth and film quality as it not only facilitates the removal of DMSO but also form homogeneous films with uniform grain sizes.[16-17] As a proof of concept, significant morphological changes using different antisolvents were observed in Figure S1. It was observed that chloroform produces the most uniform and compact films attributed to its rapid extraction of DMSO that leads to supersaturation and nucleation. Unlike other antisolvents, chloroform is non-coordinating and highly volatile, ensuring minimal interference with precursor chemistry and rapid film formation.[18] Comparatively, MeOAc leads to smaller grain sizes with moderate uniformity, whereas ethanol, IPA, and toluene result in films with voids and incomplete coverage which could be attributed to their slower extraction rates and lower miscibility with DMSO. Our morphology results are on par with the recent reports that highlight the ability of chloroform to produce high-quality films with superior morphology for optoelectronic applications.[19] We further evaluate the effect of volume of chloroform on $K_2CuBr_3$ film growth, which reveals that the 200 μL of chloroform results in uniform film with optimal grain size, while excessive chloroform causes over-saturation which causes grain aggregation (Figure S2).

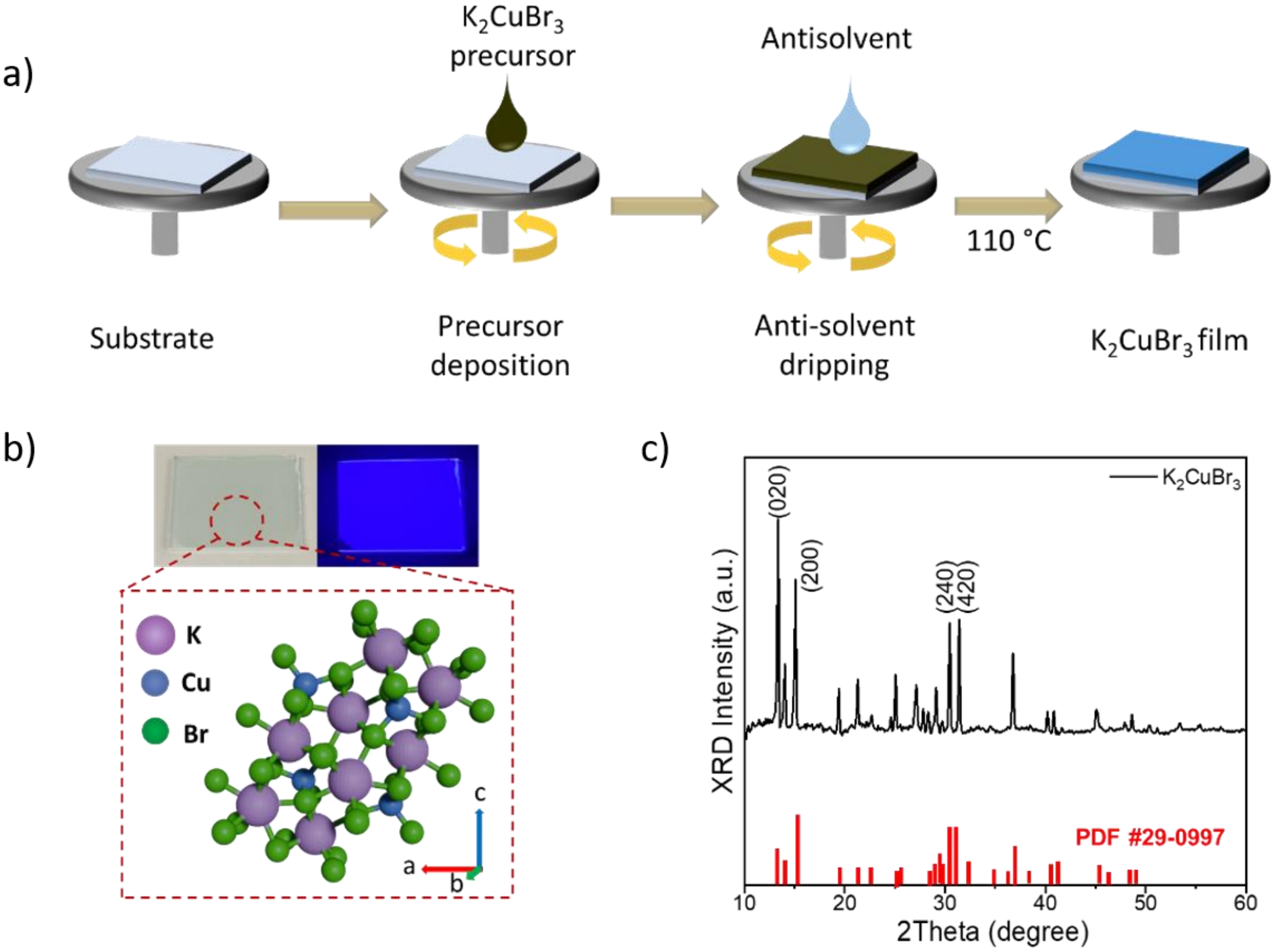

*Figure 1 Solution processed fabrication and structural characterization of thin $K_2CuBr_3$ film.* ***(a)*** *Illustration of one-step antisolvent deposition by spin coating method.* ***(b)*** *Illustration of $K_2CuBr_3$ crystal structure and photos of $K_2CuBr_3$ film under white light and UV light (342 nm).* ***(c)*** *XRD patterns of $K_2CuBr_3$ film.*

Next, we investigated the atomic structure and material properties of $K_2CuBr_3$ film. As shown in Figure 1b. One Cu atom surrounded by four Br atoms forms $[CuBr_4]$ tetrahedron orientation, sharing one corner to become $[CuBr_3]^{2-}$ and these chains are separated by $K^+$ cations. This 1D crystal structure of $K_2CuBr_3$ is expected to promote carrier confinement and strong electron-phonon coupling, which are favourable for efficient excitonic emission.[20] We successfully obtained semi-transparent blue colour and excellent fluorescence under UV light with a wavelength of 342 nm, indicating the formation of $K_2CuBr_3$ that has strong quantum confinement. The as-deposited $K_2CuBr_3$ film is further confirmed by EDX analysis, representing well distributed K, Cu, and Br on the film (Figure S3). The crystal structure of as-deposited $K_2CuBr_3$ was examined by XRD, which displays five distinct diffraction peaks at 13.48°, 14.20°, 15.21°, 30.59° and 31.55°, corresponding to the (020), (200), (120), (240), and (420) crystal planes of orthorhombic $K_2CuBr_3$ (PDF #29-0997; space group, Pnma). The chemical state of the $K_2CuBr_3$ film was further performed by X-ray photoelectron spectroscopy (XPS) measurement, as shown in Figure S4. We observed K 2p orbitals with two binding energies at 295.5 and 292.7 eV associated with K $2p_{1/2}$ and K $2p_{3/2}$, respectively, which confirms the presence of $K^+$. Analogously, Figure S4c verifies the presence of monovalent $Cu^+$ as the strong binding energies at 951.8 and 932 eV corresponding to $Cu2p_{1/2}$ and $Cu2p_{3/2}$ orbitals were detected with the energy separation of 19.8 eV. It is noted that a weak binding energy at 934.2 eV is associated with $Cu^{2+}$, which can be attributed to inevitable surface oxidation of $K_2CuBr_3$ film by air exposure. The Br 3d spectrum in Figure S4d shows two peaks at 69.7 and 68.7 eV, corresponding to the Br $3d_{3/2}$ and Br $3d_{5/2}$ orbitals. For thermal stability check, no significant weight loss was observed below 450 °C by thermogravimetric analysis (TGA), revealing that the $K_2CuBr_3$ film demonstrates superb thermal stability compared to other organic halide perovskite materials (Figure S5).

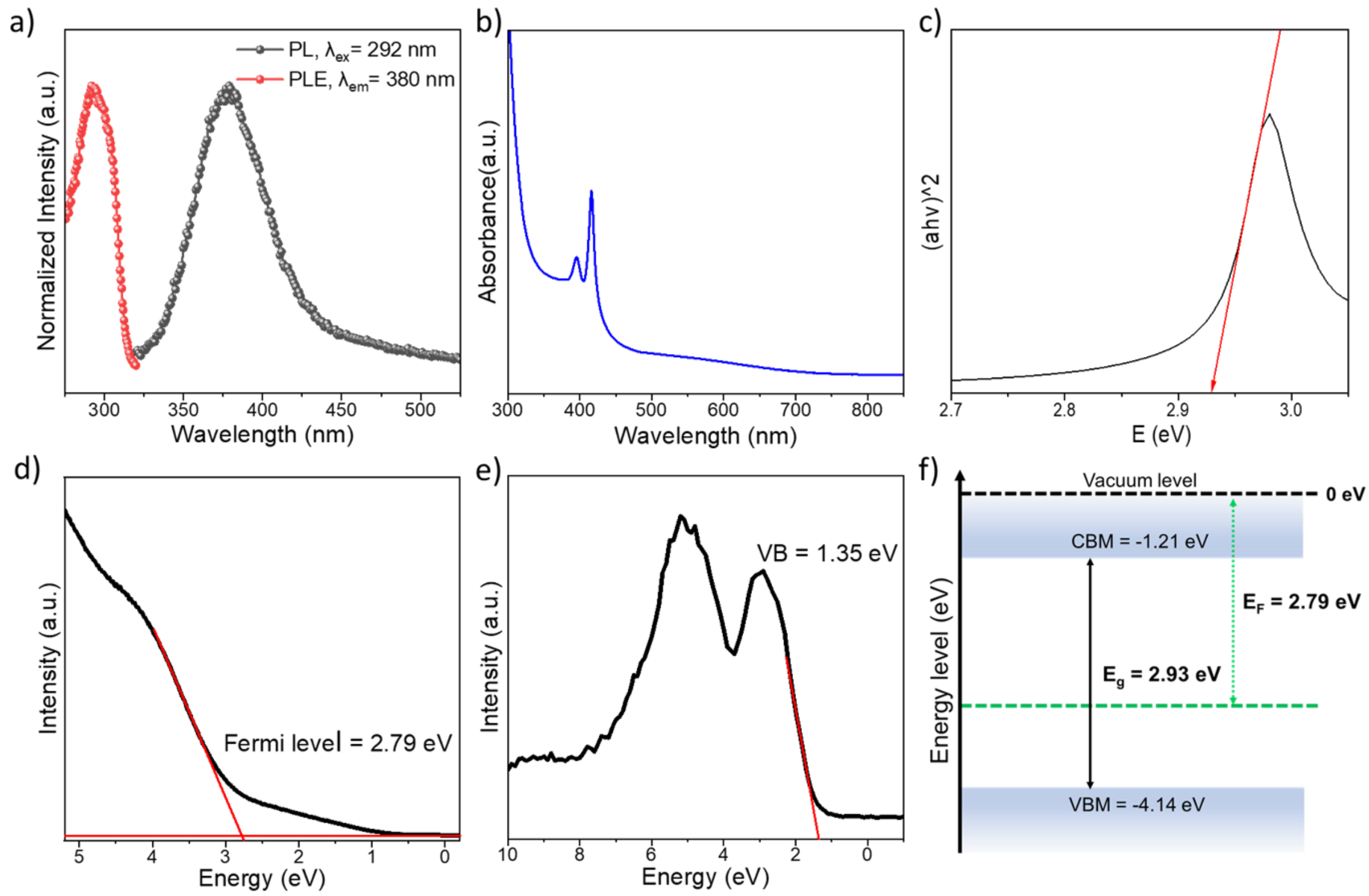

*Figure 2 Optical properties of $K_2CuBr_3$ film.* ***(a)*** *Photoluminescence excitation (PLE) and emission (PL) spectra of deposited $K_2CuBr_3$ film.* ***(b)*** *and* ***(c)*** *Ultraviolet-visible (UV-Vis) absorption spectrum of deposited $K_2CuBr_3$ film and Tauc plot based on UV-Vis absorption spectrum, respectively.* ***(d)*** *and* ***(e)*** *UPS spectrum of $K_2CuBr_3$ film.* ***(f)*** *Band diagram showing Fermi level and bandgap.*

The optical properties of $K_2CuBr_3$ film were further explored by photoluminescence spectroscopy. The photoluminescence excitation (PLE) and emission (PL) spectra for deposited $K_2CuBr_3$ film are depicted in Figure 2a. The $K_2CuBr_3$ film exhibits bright blue emission with a PL peak centred at 380 nm under UV excitation, while the corresponding PLE spectrum shows a main excitation feature at 292 nm. This emission is consistent with excitonic recombination in low-dimensional copper halides, where structural confinement and electron–phonon coupling can promote localized excited states.[21] The resulting Stokes shift in $K_2CuBr_3$ film is ~ 87 nm, which is relatively smaller than other low-dimensional perovskites. It is also noted that the full width at half-maximum (FWHM) of observed PL peaks for $K_2CuBr_3$ film are narrower compared to other STE based emitters such as copper halides ($Cs_3Cu_2Br_5$ and $Cs_3Cu_2I_5$), which indicates that $K_2CuBr_3$ are promising blue emitting material with excellent energy efficiency.

The ultraviolet-visible (UV-vis) absorption spectrum of $K_2CuBr_3$ film is shown in Figure 2b. Two pronounced absorptions are observed at 395 nm and 417 nm, suggesting strong near-UV/violet absorption associated with excitonic transitions in the low-dimensional $K_2CuBr_3$ lattice.[22,23] The direct bandgap of $K_2CuBr_3$ is estimated to be 2.93 eV using Tauc Plot as shown in Figure 2c. To further investigate the intrinsic characteristics of the $K_2CuBr_3$ film, ultraviolet photoelectron spectroscopy (UPS) measurement was performed using He I (hv= 21.2 eV) source. As shown in Figure 2d and e, we obtained the tangent of fermi level ($E_F$) and VB onset energy ($E_{VB}$) estimated at 2.79 and 1.35 eV, respectively. Given by ionization energy (IE) = work function (WF) + $E_{VB}$, IE is calculated as -4.14 eV and the fermi level is estimated to be located near VB, suggesting p-type characteristics of the $K_2CuBr_3$ film (Figure 2e).

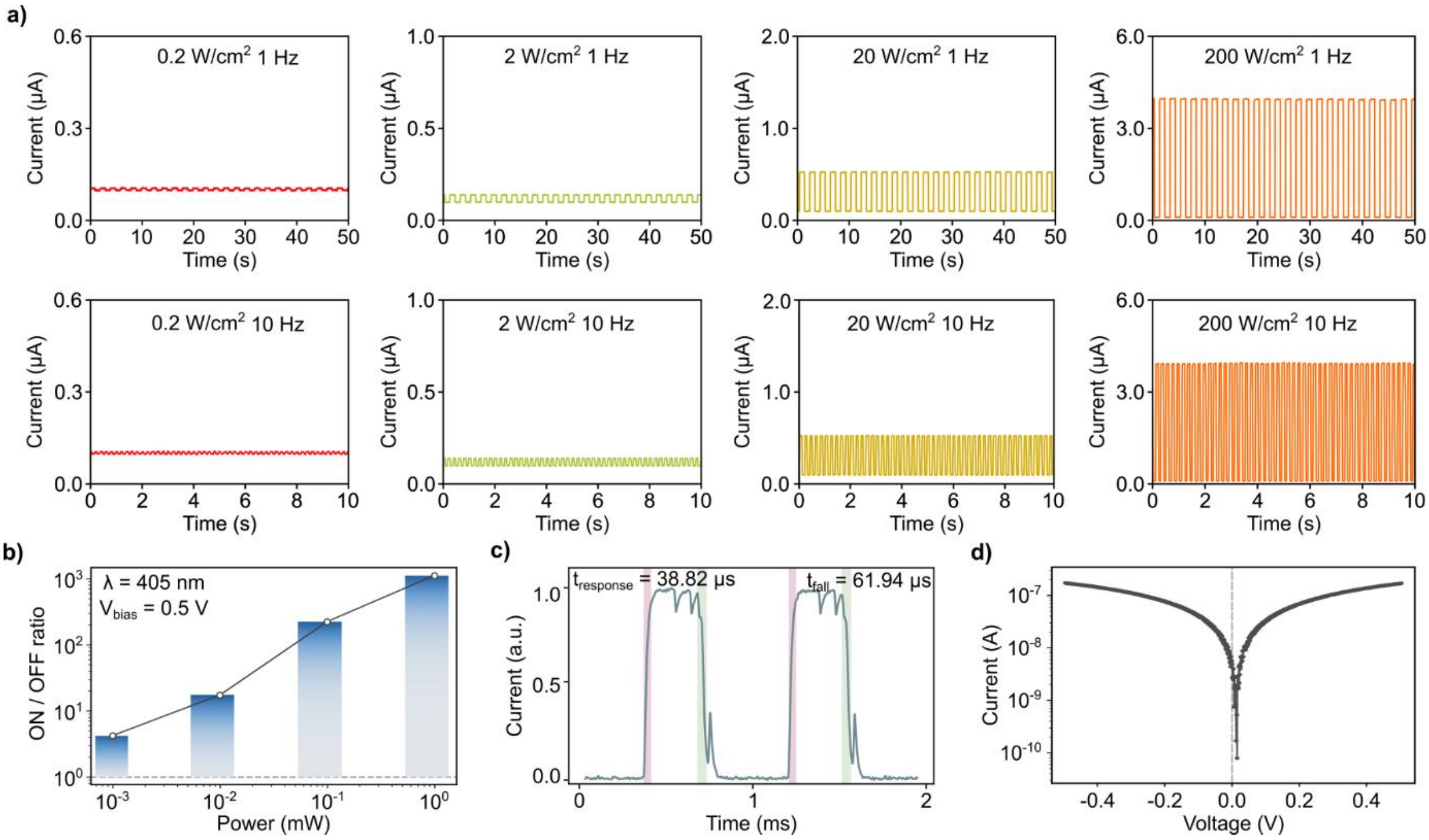

*Figure 3. Photodetection performance of the $K_2CuBr_3$ thin-film photodetector.* ***(a)*** *Time-resolved photocurrent responses of the $K_2CuBr_3$ thin-film photodetector under periodically modulated 405 nm near-UV illumination with power densities ranging from 0.2 to 200 W/cm² at modulation frequencies of 1 and 10 Hz.* ***(b)*** *Power-dependent ON/OFF ratio extracted from the temporal photocurrent response under 405 nm illumination at a fixed bias of 0.5 V.* ***(c)*** *The tested response speed of the device under 405 nm illumination, showing a rise time of 38.82 μs and a fall time of 61.94 μs.* ***(d)*** *Dark current curve measured in the absence of illumination, showing a low dark current of around $10^{-10}$ A.*

The photoresponse performance of the $K_2CuBr_3$ thin-film photodetector was evaluated using a device fabricated on platinum interdigitated electrodes. As shown in Figure 3a, the photodetector exhibits stable and reproducible photocurrent responses under periodically modulated 405 nm near-UV illumination at a bias voltage of 0.5 V. Clear photoswitching behavior was observed over a wide illumination power density ranging from 0.2 to 200 W/cm² at both 1 and 10 Hz, confirming reliable optical to electrical conversion over repeated operating cycles. The power-dependent photoresponse was further quantified by extracting the ON/OFF ratio from the photocurrent switching curves. As shown in Figure 3b, the ON/OFF ratio increases linearly with the incident optical power and reaches approximately $10^3$ at 1mW, demonstrating 3 orders of magnitude enhancement. To further evaluate the power-dependent photoresponse, the responsivity was extracted and summarized in the Methods and Figure S6. The photodetector maintains a responsivity above $3.8\times10^{-3}$ A $W^{-1}$ over the measured power range, with only a slight decrease at higher illumination intensities. This behavior suggests that the $K_2CuBr_3$ channel can provide stable photocurrent generation under different illumination powers, while the reduced responsivity at increased power may be associated with trap filling and enhanced carrier recombination.

Next, we evaluated the response and decay time of the $K_2CuBr_3$ thin-film photodetector. As shown in Figure 3c, the device exhibits a rise time of 38.82 μs and a decay time of 61.94 μs under 405 nm illumination, indicating fast carrier generation and extraction in the planar photoconductive channel. These response times are shorter than the 100 μs pattern dwell time used in our SPI measurements, suggesting that the detector can follow the 10 kHz structured illumination sequence used for image reconstruction. The power-dependent rise and fall times are further summarized in Figure S7. The dark current characteristics of the $K_2CuBr_3$ thin-film photodetector was measured in the absence of illumination. As shown in Figure 3d, the IV curve recorded from −0.5 to 0.5 V shows a low dark-current level on the order of $10^{-10}$ A. The suppressed dark current indicates low electrical noise and provides favourable conditions for weak-signal photodetection. The stable photoswitching, fast temporal response and low dark current demonstrate that the $K_2CuBr_3$ thin-film photodetector is suitable for near-UV single-pixel imaging.

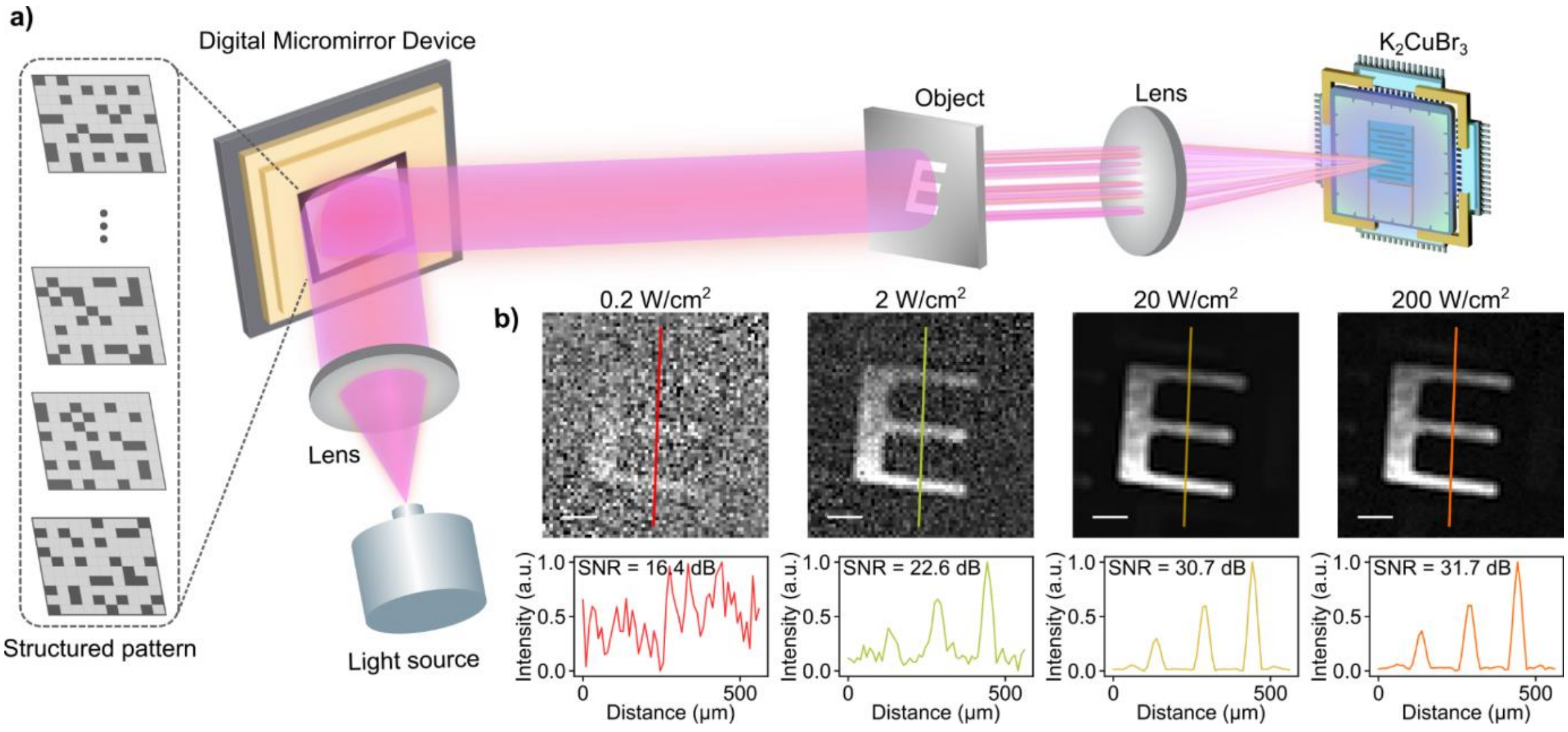

*Figure 4. Single-pixel imaging based on the $K_2CuBr_3$ photodetector.* ***(a)*** *Schematic illustration of the single-pixel imaging setup. A 405 nm light source is spatially modulated by a digital micromirror device (DMD) to generate a sequence of structured illumination patterns, which are projected onto an object mask containing the letter "E". The transmitted light is collected by a lens and focused onto the $K_2CuBr_3$ photodetector, which records the pattern-dependent integrated intensity. The object image is subsequently reconstructed by correlating the measured photocurrent signals with the corresponding illumination patterns.* ***(b)*** *Reconstructed images of the letter "E" obtained under 405 nm illumination power densities of 0.2, 2, 20, and 200 W/cm². Higher power density yields improved image contrast and*

*reconstruction fidelity, consistent with the corresponding line profiles and the increase in SNR from 16.4 to 31.7 dB. Scale bar: 100 μm.*

To evaluate the imaging capability of the $K_2CuBr_3$ thin-film photodetector, we integrated the device into a near-UV single-pixel imaging system. As illustrated in Figure 4a, the incident light was spatially encoded by a digital micromirror device and projected onto an object mask containing the letter "E". The transmitted light was collected by the $K_2CuBr_3$ thin-film photodetector, which recorded the pattern-dependent integrated photocurrent signals for image reconstruction. Details of the single-pixel imaging experimental setup are provided in the Supporting Information. Representative reconstructed images acquired under different 405 nm illumination power densities are shown in Figure 4b. At a low power density of 0.2 W $cm^{-2}$, the letter "E" remains identifiable, although the reconstructed image contains noticeable noise due to the limited photocurrent signal. As the illumination power density increases to 2, 20, and 200 W/$cm^2$, the reconstructed images exhibit progressively improved contrast and fidelity, indicating more reliable pattern-dependent photocurrent readout. This improvement is further supported by the corresponding line profiles, where the signal-to-noise ratio (SNR) increases from 16.4 to 31.7 dB. These results demonstrate the feasibility of using the $K_2CuBr_3$ thin-film photodetector as an effective single-pixel sensor for near-UV computational imaging.

The reconstructed images reveal a clear dependence of imaging fidelity on illumination irradiance. At low power density, the target pattern remains recognizable, but the image contains noticeable background fluctuation, indicating that the reconstruction is mainly limited by the magnitude and stability of the photocurrent signal. As the irradiance increases, the enhanced photocurrent response improves the contrast between illuminated and non-illuminated patterns, leading to clearer object boundaries and a higher SNR. This trend suggests that further improvement of $K_2CuBr_3$-based SPI should focus on increasing useful photocurrent while suppressing noise and temporal fluctuation. By improving film uniformity, reducing trap-assisted recombination and optimizing carrier collection are expected to enhance weak-light response and image reconstruction quality. Combining such detector optimization with low-noise readout and more efficient sampling or reconstruction strategies may further reduce the required illumination dose and acquisition time, enabling more practical lead-free UV computational imaging systems.

In summary, we have demonstrated solution-processed one-dimensional $K_2CuBr_3$ thin films as a lead-free, near-UV photoactive film and integrated into a SPI system. The optimal fabricated $K_2CuBr_3$ films exhibits a direct bandgap of 2.93 eV and strong excitonic features. The $K_2CuBr_3$ embedded planar structure photodetector demonstrated superior photoswitching across a wide irradiance range (0.2–200 W/$cm^2$) with ultrafast rise and decay time of 38.82 and 61.92 μs. Furthermore, $K_2CuBr_3$ as photoactive channel successfully reconstructs near-UV images with SNR improving from 16.4 to 31.7 dB across the full irradiance range, which confirms the compatibility of $K_2CuBr_3$ with the temporal demands of computational imaging. This work opens new opportunities for lead-free copper halide materials in compact UV computational imaging systems.

## Acknowledgements

The authors acknowledge financial support from the Australian Research Council (CE200100010, FT220100053, DP250100973, DP250100973, DE260104454, DE240100179) and the Air Force Office of Scientific Research (FA2386-25-1-4044).

## Conflicts of Interest

The authors declare no conflicts of interest.

## Data Availability Statement

The data that support the findings of this study are available from the corresponding author upon reasonable request.

Reference

(1) Li, Z.; Yan, T.; Fang, X. Low-dimensional wide-bandgap semiconductors for UV photodetectors. *Nature Reviews Materials* **2023**, *8* (9), 587-603. DOI: 10.1038/s41578-023-00583-9.

(2) Cao, F.; Liu, Y.; Liu, M.; Han, Z.; Xu, X.; Fan, Q.; Sun, B. Wide Bandgap Semiconductors for Ultraviolet Photodetectors: Approaches, Applications, and Prospects. *Research 7*, 0385. DOI: 10.34133/research.0385 (acccessed 2026/05/29).

(3) Kneissl, M.; Seong, T.-Y.; Han, J.; Amano, H. The emergence and prospects of deep-ultraviolet light-emitting diode technologies. *Nature Photonics* **2019**, *13* (4), 233-244. DOI: 10.1038/s41566-019-0359-9.

(4) Jain, N.; Kumar, D.; Bhardwaj, K.; Sharma, R. K.; Holovsky, J.; Mishra, M.; Mishra, Y. K.; Sharma, S. K. Heterostructured core-shell metal oxide-based nanobrushes for ultrafast UV photodetectors. *Materials Science and Engineering: R: Reports* **2024**, *160*, 100826. DOI: https://doi.org/10.1016/j.mser.2024.100826.

(5) Zhou, X.; Lu, Z.; Zhang, L.; Ke, Q. Wide-bandgap all-inorganic lead-free perovskites for ultraviolet photodetectors. *Nano Energy* **2023**, *117*, 108908. DOI: https://doi.org/10.1016/j.nanoen.2023.108908.

(6) Wang, Y.; Huang, K.; Fang, J.; Yan, M.; Wu, E.; Zeng, H. Mid-infrared single-pixel imaging at the single-photon level. *Nature Communications* **2023**, *14* (1), 1073. DOI: 10.1038/s41467-023-36815-3.

(7) Kilcullen, P.; Ozaki, T.; Liang, J. Compressed ultrahigh-speed single-pixel imaging by swept aggregate patterns. *Nature Communications* **2022**, *13* (1), 7879. DOI: 10.1038/s41467-022-35585-8.

(8) Meng, H.; Gao, Y.; Wang, X.; Li, X.; Wang, L.; Zhao, X.; Sun, B. Quantum dot-enabled infrared hyperspectral imaging with single-pixel detection. *Light: Science & Applications* **2024**, *13* (1), 121. DOI: 10.1038/s41377-024-01476-4.

(9) Li, W.; Hu, X.; Wu, J.; Fan, K.; Chen, B.; Zhang, C.; Hu, W.; Cao, X.; Jin, B.; Lu, Y.; et al. Dual-color terahertz spatial light modulator for single-pixel imaging. *Light: Science & Applications* **2022**, *11* (1), 191. DOI: 10.1038/s41377-022-00879-5.

(10) Yao, F.; Dong, K.; Ke, W.; Fang, G. Micro/Nano Perovskite Materials for Advanced X-ray Detection and Imaging. *ACS Nano* **2024**, *18* (8), 6095-6110. DOI: 10.1021/acsnano.3c10116.

(11) Guan, X.; Yu, X.; Periyanagounder, D.; Benzigar, M. R.; Huang, J.-K.; Lin, C.-H.; Kim, J.; Singh, S.; Hu, L.; Liu, G.; et al. Recent Progress in Short- to Long-Wave Infrared Photodetection Using 2D Materials and Heterostructures. *Advanced Optical Materials* **2021**, *9* (4), 2001708. DOI: https://doi.org/10.1002/adom.202001708 (acccessed 2026/05/29).

(12) Zhang, L.; Mei, L.; Wang, K.; Lv, Y.; Zhang, S.; Lian, Y.; Liu, X.; Ma, Z.; Xiao, G.; Liu, Q.; et al. Advances in the Application of Perovskite Materials. *Nano-Micro Letters* **2023**, *15* (1), 177. DOI: 10.1007/s40820-023-01140-3.

(13) Zhang, Z.-X.; Li, C.; Lu, Y.; Tong, X.-W.; Liang, F.-X.; Zhao, X.-Y.; Wu, D.; Xie, C.; Luo, L.-B. Sensitive Deep Ultraviolet Photodetector and Image Sensor Composed of Inorganic Lead-Free Cs3Cu2I5 Perovskite with Wide Bandgap. *The Journal of Physical Chemistry Letters* **2019**, *10* (18), 5343-5350. DOI: 10.1021/acs.jpclett.9b02390.

(14) Panchanan, S.; Dastgeer, G.; Dutta, S.; Hu, M.; Lee, S.-U.; Im, J.; Seok, S. I. Cerium-based halide perovskite derivatives: A promising alternative for lead-free narrowband UV photodetection. *Matter* **2024**, *7* (11), 3949-3969. DOI: 10.1016/j.matt.2024.07.010 (acccessed 2026/05/29).

(15) Zhou, H.; Chen, Q.; Li, G.; Luo, S.; Song, T.-b.; Duan, H.-S.; Hong, Z.; You, J.; Liu, Y.; Yang, Y. Interface engineering of highly efficient perovskite solar cells. *Science* **2014**, *345* (6196), 542-546. DOI: 10.1126/science.1254050 (acccessed 2026/05/29).
(16) Sun, J.; Li, F.; Yuan, J.; Ma, W. Advances in Metal Halide Perovskite Film Preparation: The Role of Anti-Solvent Treatment. *Small Methods* **2021**, *5* (5), 2100046. DOI: https://doi.org/10.1002/smtd.202100046 (acccessed 2026/05/29).
(17) Taylor, A. D.; Sun, Q.; Goetz, K. P.; An, Q.; Schramm, T.; Hofstetter, Y.; Litterst, M.; Paulus, F.; Vaynzof, Y. A general approach to high-efficiency perovskite solar cells by any antisolvent. *Nature Communications* **2021**, *12* (1), 1878. DOI: 10.1038/s41467-021-22049-8.
(18) Jeon, N. J.; Noh, J. H.; Kim, Y. C.; Yang, W. S.; Ryu, S.; Seok, S. I. Solvent engineering for high-performance inorganic–organic hybrid perovskite solar cells. *Nature Materials* **2014**, *13* (9), 897-903. DOI: 10.1038/nmat4014.
(19) Wang, R.; Mujahid, M.; Duan, Y.; Wang, Z.-K.; Xue, J.; Yang, Y. A Review of Perovskites Solar Cell Stability. *Advanced Functional Materials* **2019**, *29* (47), 1808843. DOI: https://doi.org/10.1002/adfm.201808843 (acccessed 2026/05/29).
(20) Gao, W.; Niu, G.; Yin, L.; Yang, B.; Yuan, J.-H.; Zhang, D.; Xue, K.-H.; Miao, X.; Hu, Q.; Du, X.; et al. One-Dimensional All-Inorganic K2CuBr3 with Violet Emission as Efficient X-ray Scintillators. *ACS Applied Electronic Materials* **2020**, *2* (7), 2242-2249. DOI: 10.1021/acsaelm.0c00414.
(21) Creason, T. D.; McWhorter, T. M.; Bell, Z.; Du, M.-H.; Saparov, B. K2CuX3 (X = Cl, Br): All-Inorganic Lead-Free Blue Emitters with Near-Unity Photoluminescence Quantum Yield. *Chemistry of Materials* **2020**, *32* (14), 6197-6205. DOI: 10.1021/acs.chemmater.0c02098.
(22) Son, J. S.; Yu, J. H.; Kwon, S. G.; Lee, J.; Joo, J.; Hyeon, T. Colloidal Synthesis of Ultrathin Two-Dimensional Semiconductor Nanocrystals. *Advanced Materials* **2011**, *23* (28), 3214-3219. DOI: https://doi.org/10.1002/adma.201101334 (acccessed 2026/05/29).
(23) Sui, X.; Gao, X.; Wu, X.; Li, C.; Yang, X.; Du, W.; Ding, Z.; Jin, S.; Wu, K.; Sum, T. C.; et al. Zone-Folded Longitudinal Acoustic Phonons Driving Self-Trapped State Emission in Colloidal CdSe Nanoplatelet Superlattices. *Nano Letters* **2021**, *21* (10), 4137-4144. DOI: 10.1021/acs.nanolett.0c04169.